# Accelerating observers measure the period of the oscillations taking place in an acoustic wave (non-longitudinal case)

Stefan Popescu[1] and Bernhard Rothenstein[2]

1) Siemens AG, Erlangen, Germany
2) Politehnica University of Timisoara, Physics Department, Timisoara, Romania

**Abstract.** *We consider a scenario that involves a stationary source of acoustic waves located at the origin of the K(XOY) inertial reference frame and a receiver that performs the hyperbolic motion at a constant altitude. The observer measures the proper reception time of successive wave crests. We investigate its dependence on the propagation speed of the wave and on the altitude at which the motion takes place.*

## 1. Introduction

In a previous paper [1] we have studied a scenario that involves a stationary source of acoustic waves and an accelerating observer who measures the period of the oscillations taking place in the wave. The moving observer performed the well known hyperbolic motion [2] taking place along the line that joins source and observer.

The scenario we propose now involves a stationary source of acoustic waves $S$ located at the origin O of its rest frame K(XOY) and an observer $R$ who performs the hyperbolic motion at a constant altitude $h$ with constant proper acceleration $g$. We could replace the source with a machine gun [3] that fires bullets in all directions in space at different emission times $t_e$. A bullet emitted at time $t_e$ hits the observer R at a reception time $t_r$, both times being displayed by the synchronized clocks of the K frame. Let $u$ be the speed of the bullets. Observer $R$ performs the hyperbolic motion

$$x = \frac{c^2}{g}\cosh\frac{g}{c}t' \qquad \text{respectively} \qquad x = cT\cosh\frac{t'}{T} \qquad (1)$$

where $g$ represents the constant acceleration of the moving receiver and $t'$ represents the time displayed by his wrist watch. In the right side equation we use the notation $T = \frac{c}{g}$ in order to simplify the equations and to better underline the physical significance of various terms. Here $T$ may be interpreted as a "magic" time interval required accelerating an object with constant acceleration $g$ until it reaches the speed $c = gT$ in a non-relativistic approach. Let $t$ be the time displayed by a clock of the K frame when $R$ is located in front of it his clock reading *t'*. The two clock readings are related by

$$t = \frac{c}{g}\sinh\frac{g}{c}t' \qquad \text{respectively} \qquad t = T\sinh\frac{t'}{T} \qquad (2)$$

When represented in the world coordinates [*t, x/c*] equations (1) and (2) correspond to the parametric equations of a conjugate hyperbola. Accordingly this motion is also known as the "hyperbolic motion". It begins at $x = +\infty$, $t = -\infty$ with $V = c$ when the decelerating object approaches the origin until it reaches the rest at $t = 0$ and $x_0 = c^2/g$. Thereafter for $t > 0$ the moving object reverse direction receding the origin and accelerating toward $V = c$ as $x \rightarrow +\infty$, $t \rightarrow +\infty$ in accordance with the requirements of special relativity. A similar scenario that involves a light source was studied by Neutze and Moreau [4].

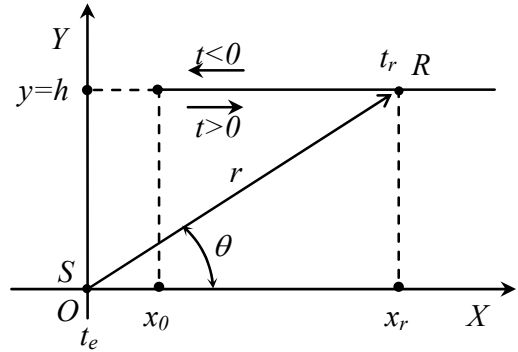

**Figure 1**. The stationary source located at the origin O and an accelerating receiver maintaining constant altitude *y=h*

Figure 1 illustrates the scenario we follow at time $t_r$ ($t'_r$) when R receives the bullet previously fired at time $t_e$. The invariance of distances measured perpendicular to the direction of relative motion requires that

$$y = y'$$

and because the bullet travels a distance

$$u(t_r - t_e)$$

Pythagoras' theorem applied to Figure 1 leads to

$$h^2 + x_r^2 = u^2(t_r - t_e)^2. \qquad (3)$$

Expressing in (3) $x_r$ and $t_r$ as a function of the time measured by *R* we have

$$h^2 + \left(cT\cosh\frac{t'_r}{T}\right)^2 = u^2\left(T\sinh\frac{t'_r}{T} - t_e\right)^2 \qquad (4)$$

Solved for $t'_r$ (4) leads to

$$t'_{r\pm} = T \arg\sinh\left(\frac{t_e \pm \beta^{-1}\sqrt{t_e^2 + (1-\beta^{-2})(T_h^2 + T^2)}}{(1-\beta^{-2}) \cdot T}\right) \tag{5}$$

where $T=c/g$, $T_h=h/u$ and $\beta = u/c$. Here $T_h$ is the time interval required for the bullets to reach for the first time the altitude $h$ of the receiver trajectory. We illustrate in figure 2 the variation of the two mathematical solutions of (5) as function of the emission time $t_e$ for $\beta = 0.2$ $T_h = 5$ and $T = 1$. This representation reveals that it is difficult to select the physically valid solution respectively the physically valid range for the mathematical solutions of (5).

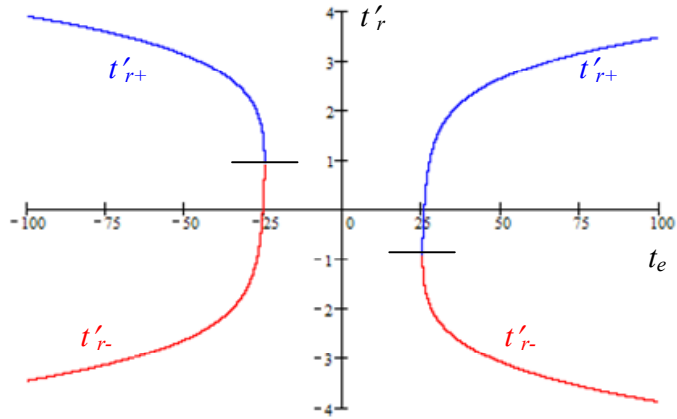

**Figure 2**. The two mathematical solutions for the reception time depending on the sign in front of the square root

To solve this problem we transpose it in a special world diagram, that we previously elaborated and used for similar purposes in [1]. For this we calculate and represent the world line of the virtual point $W(x_w,t)$ representing the intersection between the circular wave-front of the propagating acoustic signal and the receiver trajectory at $y = h$.

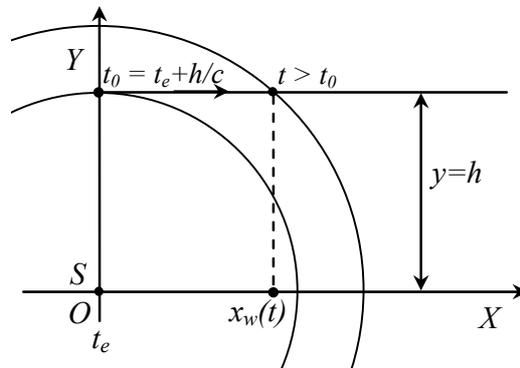

**Figure 3**. The intersection between the circular wave-front of the acoustic signal and the trajectory of the accelerating receiver at altitude $y=h$

Pythagoras' theorem applied to Figure 3 leads to
$$x_w^2 + h^2 = u^2(t-t_e)^2. \qquad (6)$$
which solved for $x_w$ returns the horizontal position of the $W$ point as:
$$x_w(t) = \sqrt{u^2(t-t_e)^2 - h^2} \qquad (7)$$
and its speed:
$$v_w(t) = \frac{u}{\sqrt{1 - \frac{h^2}{u^2(t-t_e)^2}}} \qquad (8)$$
Equation (6) describes a hyperbola. We illustrate both functions in Figure 4.

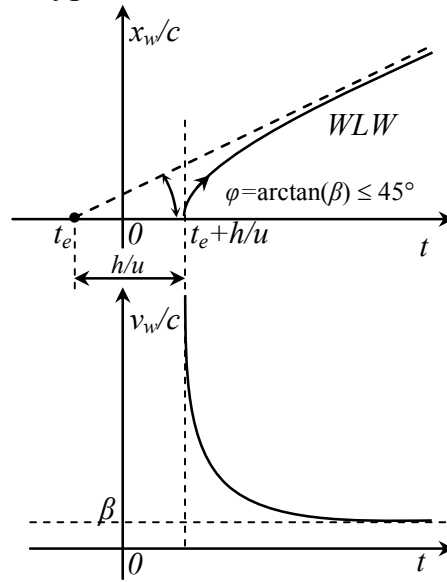

**Figure 4**. The world line and the speed of the intersection point between the wave-front of the acoustic signal and the receiver trajectory

Further conclusions result after depicting the world line of the accelerating receiver (*WLR*) and the word line of the *W* point (*WLW*) in a common world diagram. Figure 5 provides valuable information for selecting and limiting the range of the physically valid solutions. We depicted two emission-receptions events. A first conclusion is that the reception is possible only if the emission time $t_e$ is negative, which means that the receiver was approaching the source at the emission time. If the source emits at $t_e > 0$ when the receiver is receding the source then the reception is not possible at all. Therefore we shall eliminate the right side of the plots in figure 2; only the left side plots are physically valid.

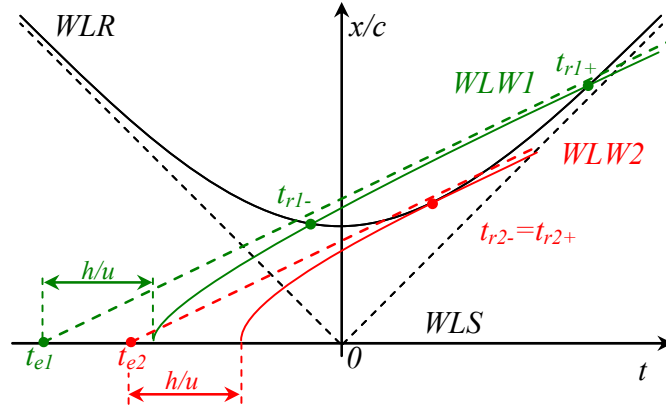

**Figure 5**. The world diagram for different emission times. A reception is possible only for early emissions taking place in the approaching phase.

Furthermore, when following the world lines in figure 5 we distinguish the following characteristic situations depending on the values of *u/c* and emission time:

- WLW doesn't intersect WLR at all and thus no information exchange between source and receiver takes place,
- WLW is tangent to WLR ($t'_{r+} = t'_{r-}$) characterized by the emission time

$$t_{e-last} = -\sqrt{(\beta^{-2}-1)(T_h^2+T^2)} \qquad (9)$$

and the reception time when R receives the last signal

$$t'_{r+} = t'_{r-} = T \operatorname{arg sinh} \sqrt{\frac{T_h^2+T^2}{(\beta^{-2}-1)\cdot T^2}} \qquad (10)$$

As expected with $u \to c$ we have $t_{e-last} \to 0$ and $t'_{r\pm} \to +\infty$. The last emission and reception times for which a communication is still possible do depend on the receiver altitude.

- WLW intersects WLR twice at the times $t'_{r-}$ and $t'_{r+}$ respectively. In this case both solutions given by (5) are valid, which means that the acoustic wave-front intercepts the accelerating receiver twice. If the stationary source emits periodical signals at time instants $t_{e,N} = N \cdot T_e$ then the approaching observer R receives these signals (bullets or the wave crests) in the succession –N,-(N-1)...-N$_{last}$, at times $t'_{r-,N}$ and thereafter he receives the same signals ones again in the reversed order at times $t'_{r+,N}$.

Now that we identified the physically valid solutions of equation (5) and their validity range we depict in figure 6 the variation of reception times with the emission time $t_e$ for different values of $T_h = h/c$.

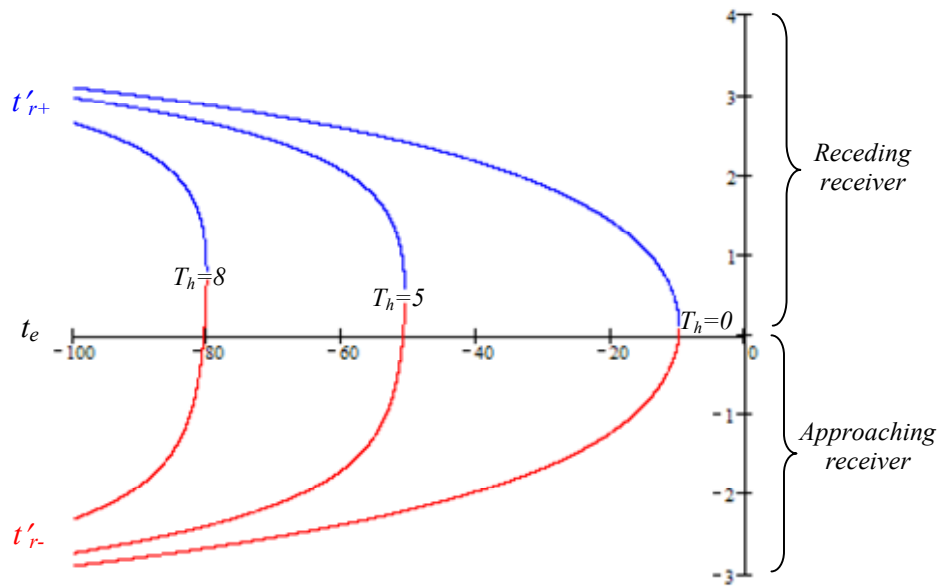

**Figure 6**. The variation of reception times with the emission time for $T = 1$, $\beta = 0.1$ and for different values of $T_h$ mentioned on the corresponding curves

Figure 7 illustrates the influence of $\beta = u/c$.

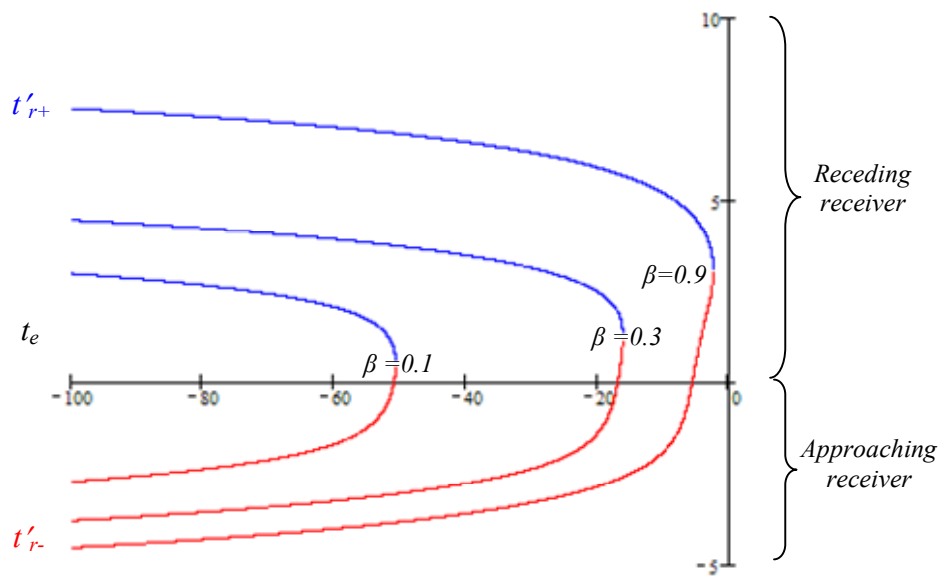

**Figure 7**. The variation of reception times with the emission time for $T_h = 5$, $T = 1$ and for different values of $\beta$

If the source emits periodically very short audio signals at time instants $t_{e,N} = N \cdot T_e$ then the accelerating observer receives the $N^{th}$ light signal when his wrist watch reads $t'_{r,N}$. The nonlinearities we notice in figure 6 and 7 suggest that the proper emission period and with it the Doppler factor defined as the quotient of proper reception and proper emission periods will reveal some oscillations similar to the one we found in the optic case [5]. These oscillations become further significant as *h* or *β* increases.

**6. Conclusions**

We have revealed some peculiarities of the acoustic Doppler Effect in the non-longitudinal case. In order to be more intuitive we have shown that we can replace the source of acoustic waves with a machine gun that fires successive bullets in all directions in space. Our approach illustrates the non-locality in the period measurement by an accelerating observer.

From a pedagogical point of view our approach teaches the student how to put in equations a given physical scenario and how to distinguish between the solutions with physical meaning, encouraging the use of computer in order to visualise the derived solutions.